\documentclass[twocolumn,showpacs,preprintnumbers,amsmath,amssymb]{revtex4}

\usepackage{graphicx,color}
\begin{document}

\addtolength{\topmargin}{2cm}

\title{Collapse of Langmuir solitons in inhomogeneous plasmas}

\author{Y.A.Chen, Y.Nishimura, and Y.Nishida}
\affiliation{Institute of Space and Plasma Sciences \\
National Cheng Kung University, Tainan 70101, Taiwan}
\email{nishimura@pssc.ncku.edu.tw}
\author{C.Z.Cheng}
\address{Graduate School of Frontier Sciences \\
University of Tokyo, Chiba 277-8561, Japan}

\begin{abstract}
Propagation of Langmuir solitons in inhomogeneous plasmas is investigated numerically.
Through numerical simulation solving Zakharov equations, 
the solitons are accelerated toward the low density side. 
As a consequence, isolated cavities moving
at ion sound velocities are emitted.
When the acceleration is further increased,
solitons collapse and the cavities separate into two lumps released
at ion sound velocities.
The threshold is estimated by an analogy between the soliton and a particle
overcoming the self-generated potential well.

\end{abstract}

\maketitle

Langmuir solitons were theoretically predicted by Zakharov \cite{zak72}.
In a homogeneous medium, the Langmuir solitons can propagate without changing their form when 
the non-linearity from a ponderomotive force and dispersion balance exactly.
Following the theoretical prediction, Langmuir solitons were observed in
a laboratory experiment
via the trapping of density cavities by imposing an external radio frequency (RF) electric field \cite{kim74}.
In the experiment, the external electromagnetic waves underwent  
mode conversion to become electrostatic waves \cite{ste74}.
It should be noted that the Langmuir soliton is also referred to as a caviton, 
and consists of one electric field soliton and one density cavity.

Seen from a practical application point of view, the generation and collapse of
Langmuir solitons are closely related to space weather forecasts.
Langmuir turbulence driven by oscillating two stream instabilities
is caused by the acceleration of electrons via the localized electric field 
of Langmuir solitons \cite{nis68,mor74}.
More specifically, type III solar bursts induce Langmuir turbulence \cite{gol84},
which in turn emit radio waves with frequencies at higher harmonics that reach 
earth in eight minutes and act as precursors for geomagnetic storms.
Such storms typically arrive a few days later. 
Understanding Langmuir solitons and Langmuir turbulence        
is one of the central issues for the space weather forecast program.

In this paper, we first investigate whether Langmuir solitons can stably propagate in inhomogeneous media.
We then examine the collapse mechanism of the solitons and discuss the threshold for the collapse.
The acceleration of Langmuir solitons in inhomogeneous media is studied analytically
employing the nonlinear Schr\"{o}dinger 
equation (NLSE) \cite{che76} and Zakharov equations \cite{chu77}
in a small acceleration limit.
Nevertheless, the dynamics of the solitons at large acceleration and their survival threshold
remains a question to be clarified.
In this work, we employ a set of nonlinear fluid equations,
the Zakharov equations to investigate the acceleration of Langmuir solitons with inhomogeneous plasma background 
densities by numerical simulation.
Based on the findings, we introduce the idea of quasi-particles, through which we interpret the acceleration as
an analogy of a point mass falling down the potential well in comparison with the Shr\"{o}dinger equation
in quantum mechanics.

As elaborated later,
we discovered a by-product of the soliton acceleration, 
namely the emission of density cavities \cite{kaw82} observed propagating
exactly at ion sound velocity. 
When the density gradient is large enough to accelerate the electric field soliton 
(and thus gain kinetic energy) to overcome a potential energy well 
produced by the density cavity, the soliton can completely escape from the cavity.
The density cavities will then remain as lumps without any sustaining mechanism
and split into pulses propagating at the ion sound velocity.

\begin{figure}[tb]
\centering
\includegraphics[height=5.6cm,angle=+00] {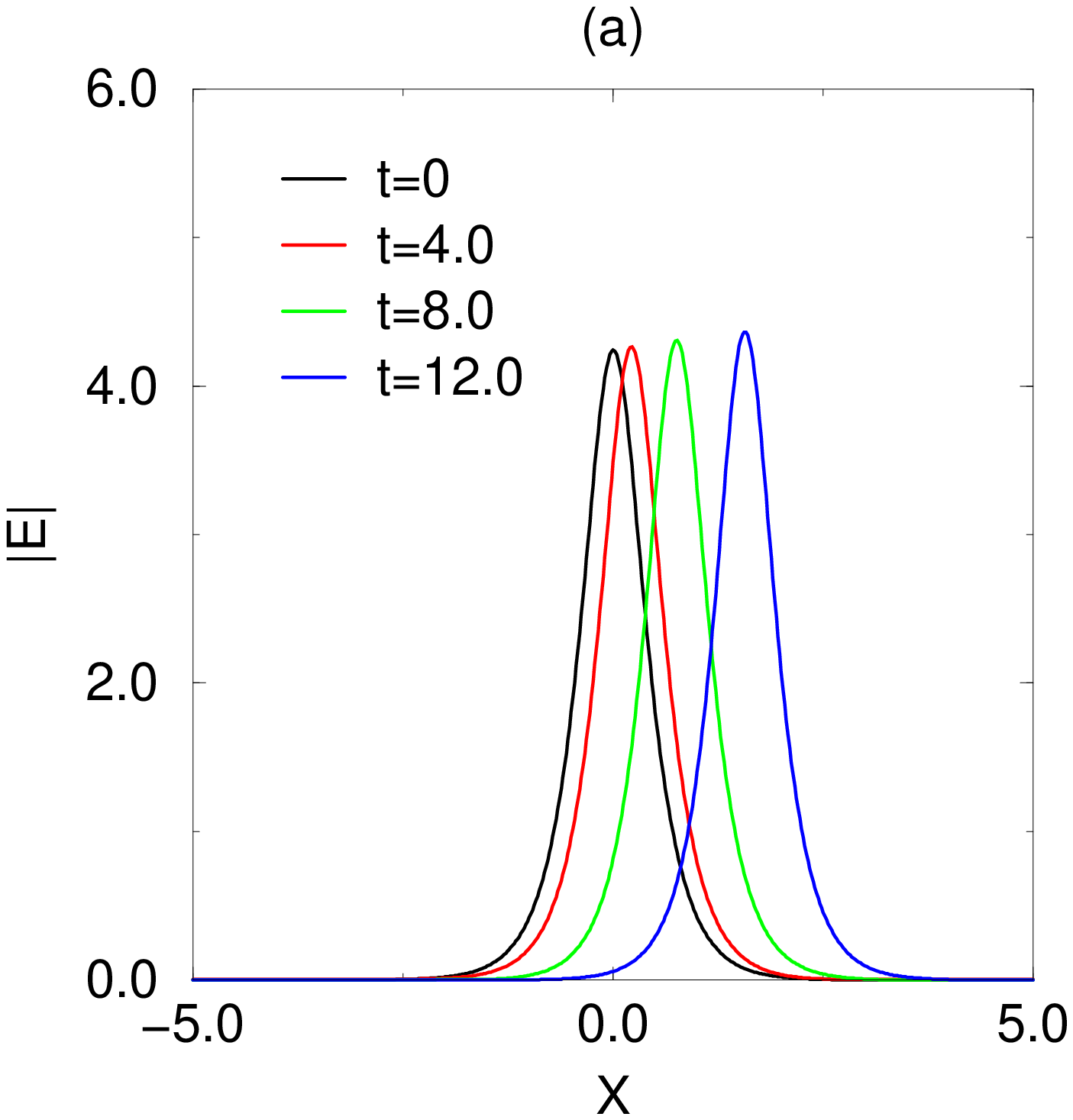}
\includegraphics[height=5.6cm,angle=+00] {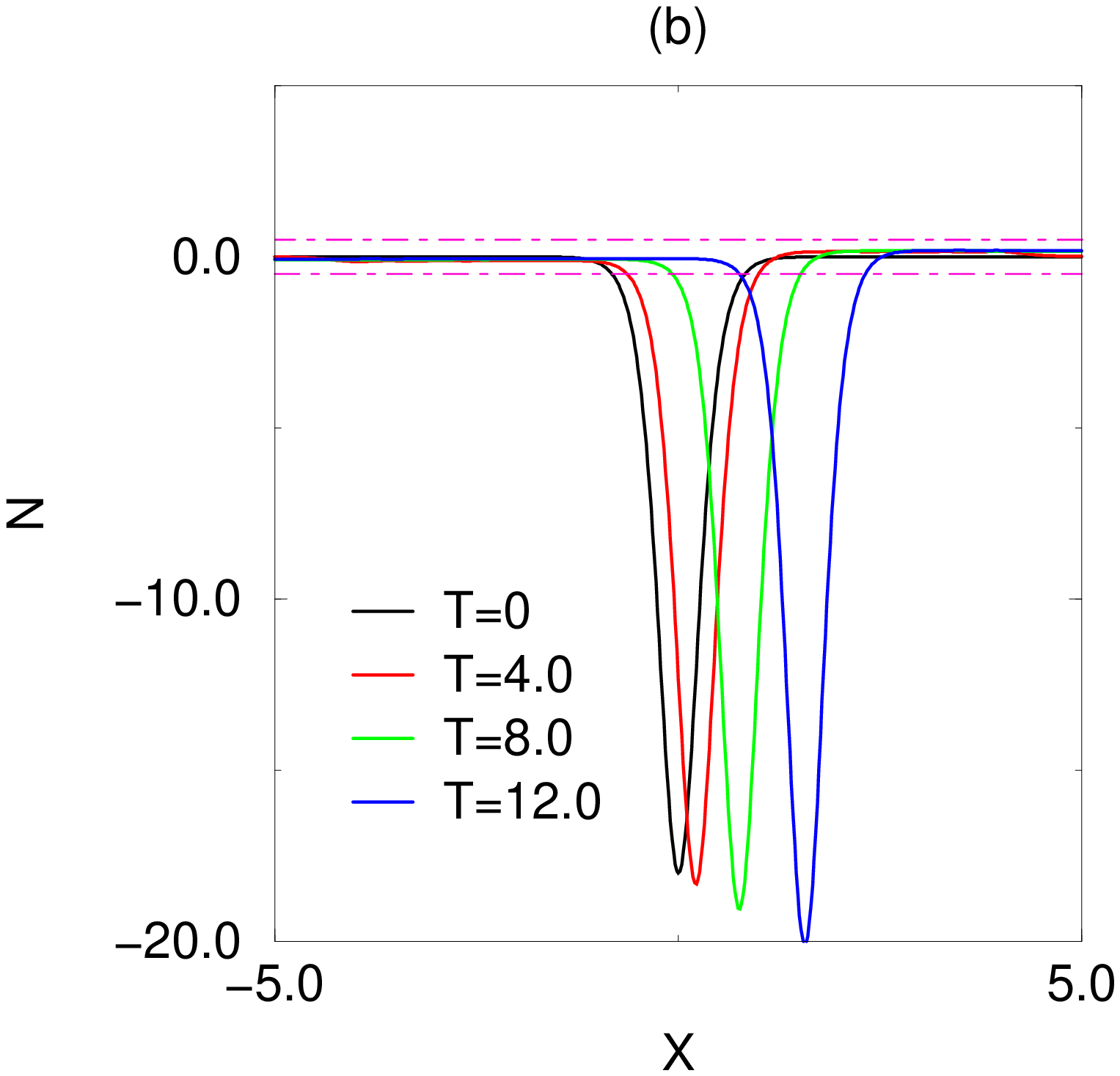}
\includegraphics[height=5.6cm,angle=+00] {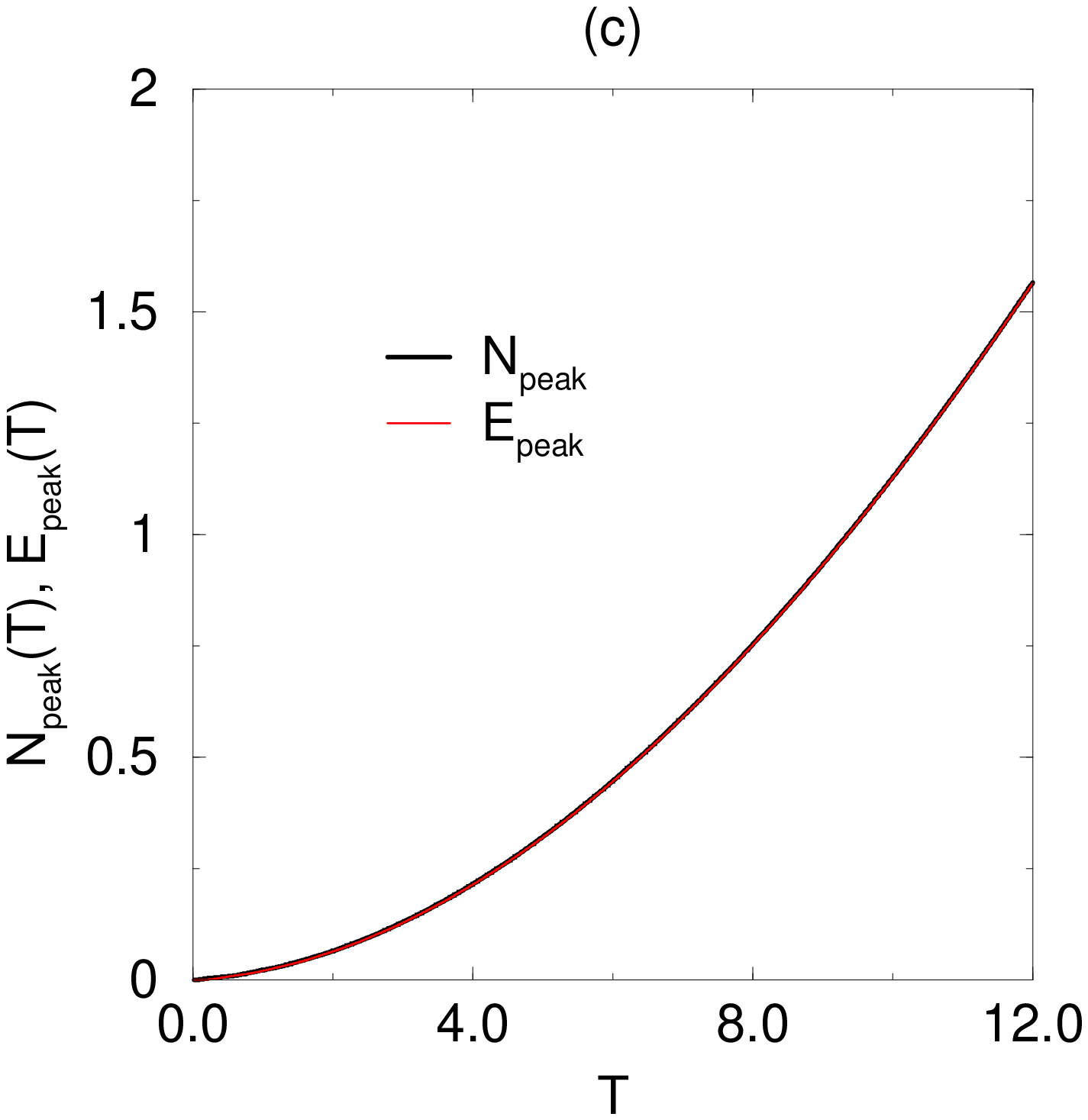}
\caption{(a) Electric field soliton, and (b) density cavity at $L = 5 \times 10^3$.
(c) Time versus the positions of the electric field peak and the density peak.}
\end{figure}

\begin{figure}[tb]
\centering
\includegraphics[height=5.6cm,angle=+00] {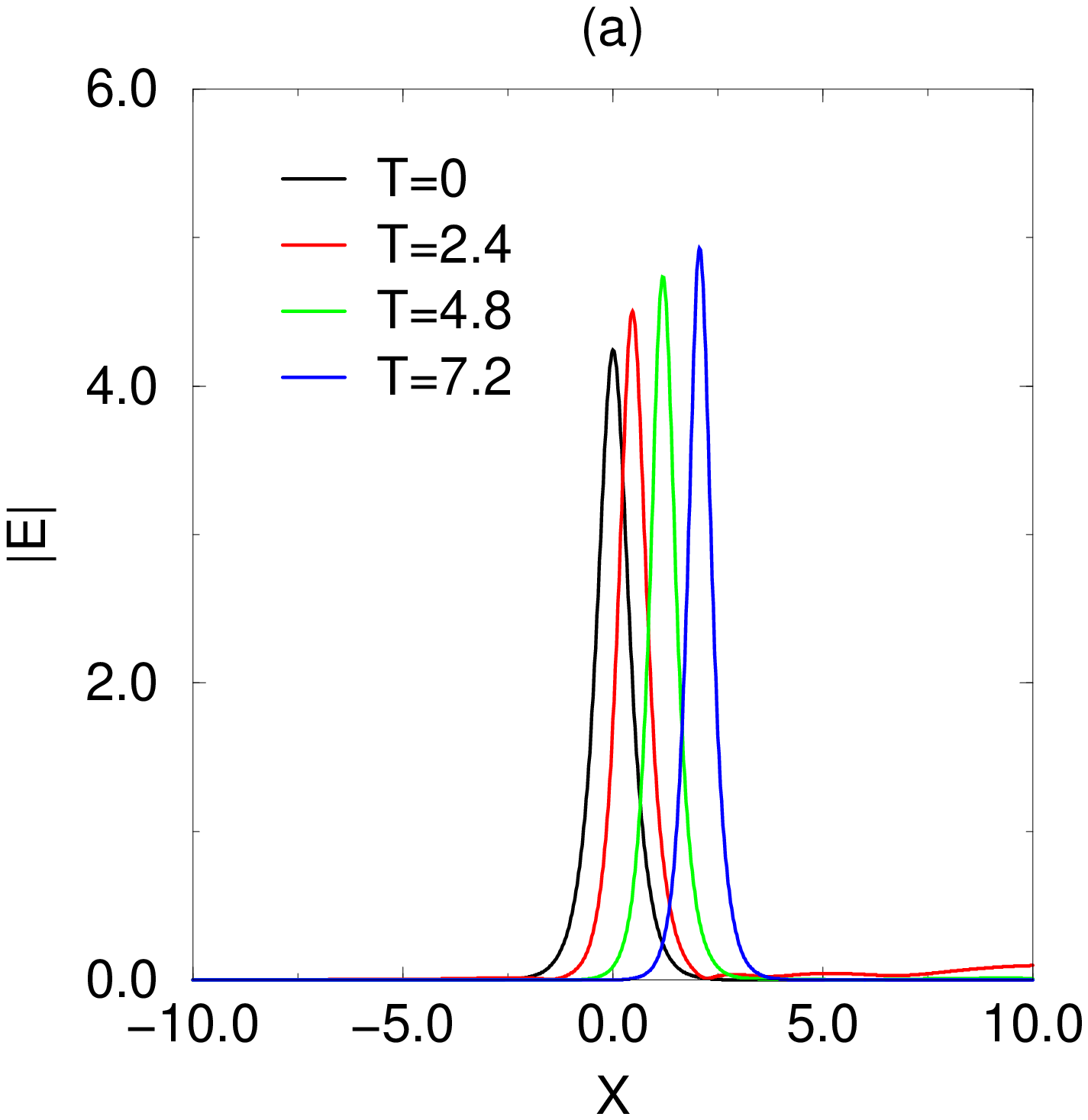}
\includegraphics[height=5.6cm,angle=+00] {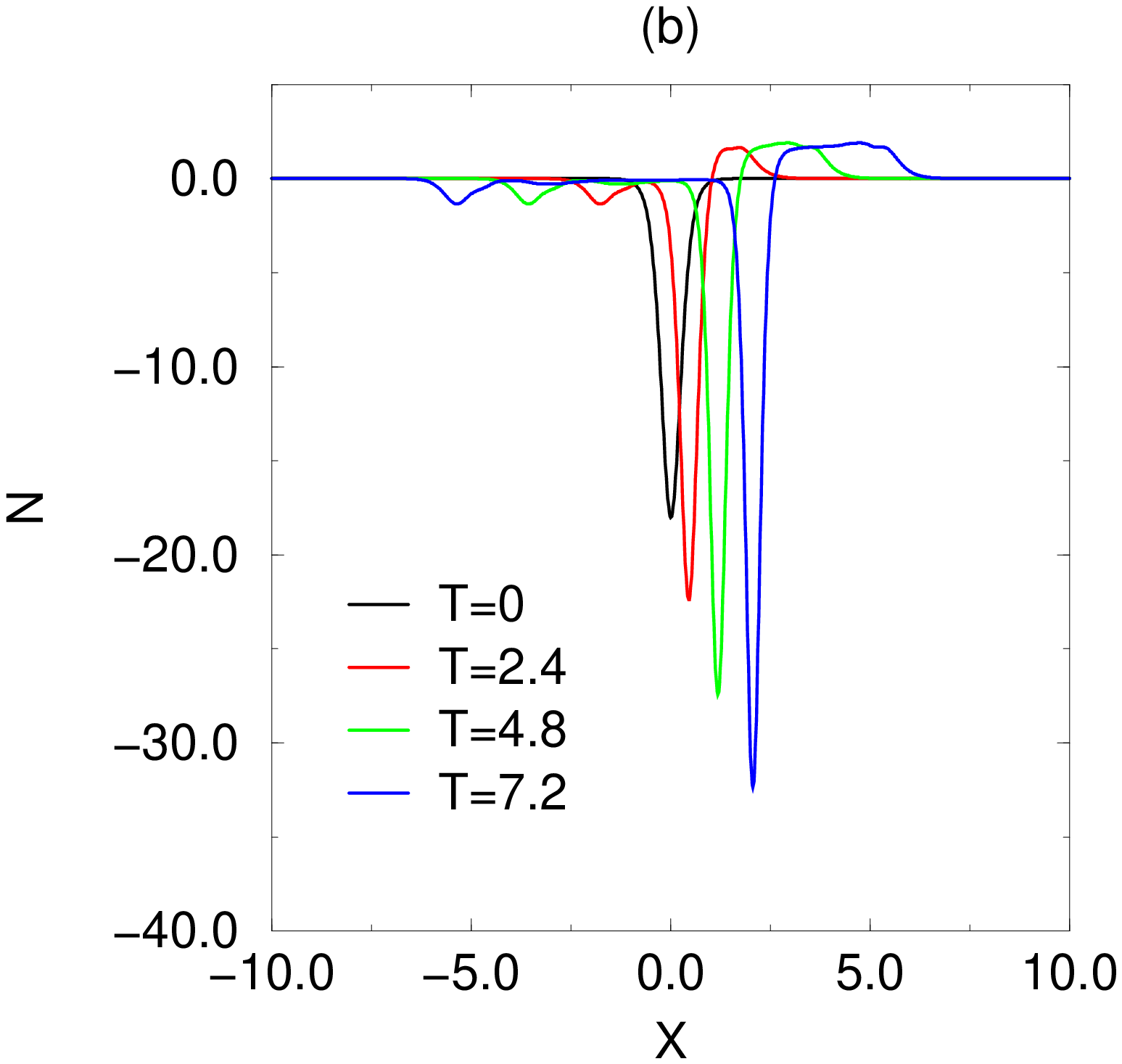}
\includegraphics[height=5.6cm,angle=+00] {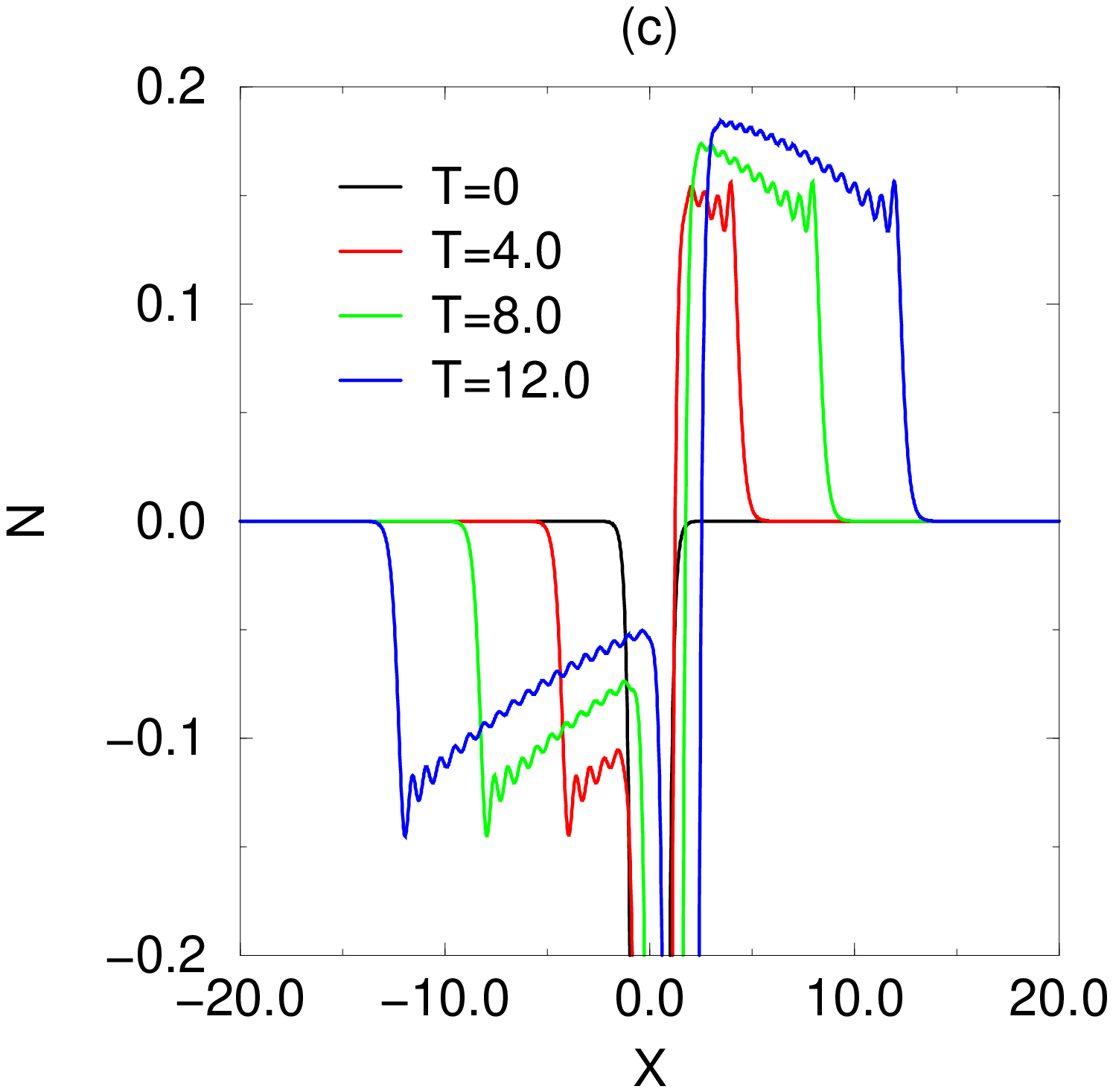}
\caption{(a) Electric field soliton, and (b) Density cavity at $L = 500$.
(c) An expansion of Fig.1 (b) corresponding to the boxed region signified by
the dashed lines.}
\end{figure}

As mentioned, we solve the Zakharov equations numerically, in this work.
Normalized Zakharov equations \cite{zak72,nic92} are given by
\begin{eqnarray}
i \partial_T E + \partial_X^2 E = N E 
\label{zak1}
\end{eqnarray}
\begin{eqnarray}
\partial^2_T N - \partial_X^2 N =  \partial_X^2 |E|^2
\label{zak2}
\end{eqnarray}
where $E$ is the slowly varying part of the electric field and $N$ is the plasma density
(ion and electron densities are connected through quasi-neutrality).
Note that all the variables in Eqs.(\ref{zak1}) and (\ref{zak2}) 
are normalized by $3 M / 2 \omega_e$ for time $T$ and $3 M^{1/2} \lambda_e /2$ for space $X$
(here $ M =m_i/m_e$ is the mass ratio where $m_i$ and $m_e$ are the ion and the electron mass, respectively), 
but not by a conventional plasma frequency ($\omega_e$) or the Debye length ($\lambda_e$)
to ensure all terms are at an order of unity. 
Equation (\ref{zak1}) is derived from electron fluid equations and the Poisson's equation,
and represents the Langmuir wave dynamics with the modulation by density change on the right side.
Equation (\ref{zak2}) is essentially the ion acoustic wave equation 
with the ponderomotive force on the right entering through quasi-neutrality. 
The NLSE \cite{gin50} can be obtained by
assuming a steady state density and thus inputting $N=-|E|^2$ into Eq.(\ref{zak1}).
The balance between the nonlinear drive and wave dispersion in the NLSE sustains the soliton.

The inhomogeneous density enters through the change in background plasma frequency in
the electron fluid equation \cite{chu77}. 
Let us now consider the derivation of the electron Zakharov equation
to elucidate the origin of where the inhomogeneous background density enters. 
In a one-dimensional form \cite{nic92}, we have 
\begin{eqnarray*}
\partial_t^2 E_h - 3 v_e^2 \partial_x^2 E_h + \omega_e^2 \left[ \frac{n_{eq} (x)}{n_0} \right] E_h 
= - \omega_e^2 \left[ \frac{n_{l}}{n_0} \right] E_h 
\end{eqnarray*}
where $v_e = \lambda_e \omega_e$ is the electron thermal velocity and 
$ n_{eq}(x)$ is the equilibrium density profile 
(the space and time variables $x$ and $t$ have dimensions).
The first and third terms eliminate each other in a homogeneous plasma
by setting a rapidly varying part of the electric field as $E_h = (1/2) \tilde{E} e^{-i \omega_e t} + c.c.$. 
Here, $n_l$ is a slowly varying part of the density perturbation.
If we assume a linear form $ n_{eq} (x) = n_0 (1 - x/l)$, we then have
\begin{eqnarray}
i \partial_T E + \partial_X^2 E =  - \alpha X E + N E  
\label{zak4}
\end{eqnarray}
for the electron equation \cite{pet61,che76,chu77}.
The acceleration parameter is given by $\alpha = 3 M/4 L$ where $L = (2  M^{-1/2} / 3 \lambda_e) l $.
We employ a hydrogen to electron mass ratio $M=1836$ for numerical computations in this paper. 
Note that the volume integral of the electric field pressure and plasma pressure,
that are the values of $\int_{-\infty}^{\infty} E^2 dX$ and $\int_{-\infty}^{\infty} N dX$ respectively, 
are invariants even when the solitons are accelerated.

Equations (\ref{zak2}) and (\ref{zak4}) are time advanced numerically by finite difference method
(the leapfrog scheme, instead of conventionally employed split step Fourier methods \cite{tah84}) 
so that non-periodic boundary conditions can be incorporated.
The analytical solution is taken as the initial condition \cite{per77}.
Note that Langmuir soliton solutions are parameterized by two free parameters, namely $K_0$ and $K_1$. 
General solutions for Eqs.(2) and (3) are given by \cite{per77}
\begin{eqnarray}
E (X,T) = E_0 \cdot sech \left[ K_0 \left( X - V_g T \right) \right] 
e^{- i \left[ K_1 X - \left( {K_1}^2 -{K_0}^2 \right) T \right] }
\label{sol1}
\end{eqnarray}
\begin{eqnarray}
N (X,T) = - 2 {K_0}^2 \cdot sech^2 \left[ K_0 \left( X - V_g T \right) \right] .
\label{sol2}
\end{eqnarray}
Here, $ {E_0}^2 =  2 {K_0}^2 \left( 1 - V_g^2 \right) $, while $V_g = 2 K_1$ is the group velocity.
The parameters taken are $K_0=3$ and $K_1=0$.
We start from $V_g=0$ to observe the pure acceleration mechanism. 

The first numerical simulation result presented in Fig.1 is 
for a Langmuir soliton with a background density change given by $L = 5 \times 10^3$.
Figures 1 (a) and (b) demonstrate acceleration of the electric field soliton and the density cavity.
while Fig.1(c) shows the trace of the peak position of the density cavity measured from Figs.1(a) and (b)
suggesting nearly constant acceleration in the initial phase
when the solitons are located inside the inhomogeneous plasma.
The peak positions are estimated by a quadratic fitting 
of the solitons and the cavities [the fitting is not in Fig.1(c) but within the $X$ space],
because the spatial resolution is restricted by the mesh size of the finite difference method.
The quadratic fitting allows us to accurately estimate the velocity and the acceleration of the solitons.
We see a parabolic increase of the soliton peak position in time, which suggests
almost constant acceleration.
Since the density gradient is small, the soliton position evolves according to $X(T) = A T^2/2$
(the rate of the acceleration is $A = 2 \alpha = 0.54 $).
Note that the velocity of the soliton itself is still sub-sonic.

In the second case, presented in Fig.2, 
we consider a shorter density scale length (meaning larger acceleration) compared to Fig.1.
Here, $L = 500$ is taken.
The emission of density cavities moving exactly at the ion sound velocity is observed,
the direction of which is opposite to the (electric field) soliton acceleration. 
On the other hand, the emission of the positive density perturbation is along the direction
of the soliton acceleration.
Subsequent to the emission discovery, we reexamined the $L = 5 \times 10^3$ case and confirmed that
the emitted density cavities (traveling at the ion sound velocity) 
were there, as long as the acceleration was finite.
Figure 2(c) shows the expansion of Fig. 1(b) (as we discuss later, small ripples
correspond to bounce motions of the soliton within the cavity).
It should be noted that the ion sound velocity is unity in our unit \cite{zak72}.
In both cases (Figs.1 and 2), the values of $\int_{-\infty}^{\infty} E^2 dX$ and $\int_{-\infty}^{\infty} N dX$ still conserve
to the relative error of $\le 10^{-5}$.
Note that isolated density cavities can be generated from any imbalance between solitons and cavities.
For example, if we set mismatched initial conditions in Eqs.(4) and (5)
(i.e. whenever the balance between the electric field pressure and 
plasma pressure by the density cavity breaks),
one can still observe cavity emission.
The plasma inhomogeneity, however, is one of the ubiquitous ingredients of the balance breaking.

We now take the shorter density scale length of $L = 50$. For a plasma with an electron temperature of
$1 eV$ and density of $10^{10} cm^{-3}$, $L=50$ corresponds to $24cm$
(the soliton width is approximately $1 cm$ with $K_0=3.0$).
As a reference,
with a further increase in the density gradient, we observe the collapse of the solitons.
The electric field quickly spread with the density inhomogeneity above the threshold
and the density cavity loses its sustaining mechanism by the ponderomotive force.
In the absence of the sustaining electric field (ponderomotive force),
the original density cavity splits into two lumps
(which then propagate in opposite directions at the ion sound velocity). 
Note that the density lumps are both negative 
which is in contrast to the subsonic acceleration cases of Figs.1 and 2
(one being a cavity, the other being a positive density perturbation).
Here in Fig.3, a total of $4 \times 10^3$ mesh points within $-40 \le X\le  60$ were prepared 
beforehand to incorporate the rapid spread of the electric field.
\begin{figure}[tb]
\centering
\includegraphics[height=5.6cm,angle=+00] {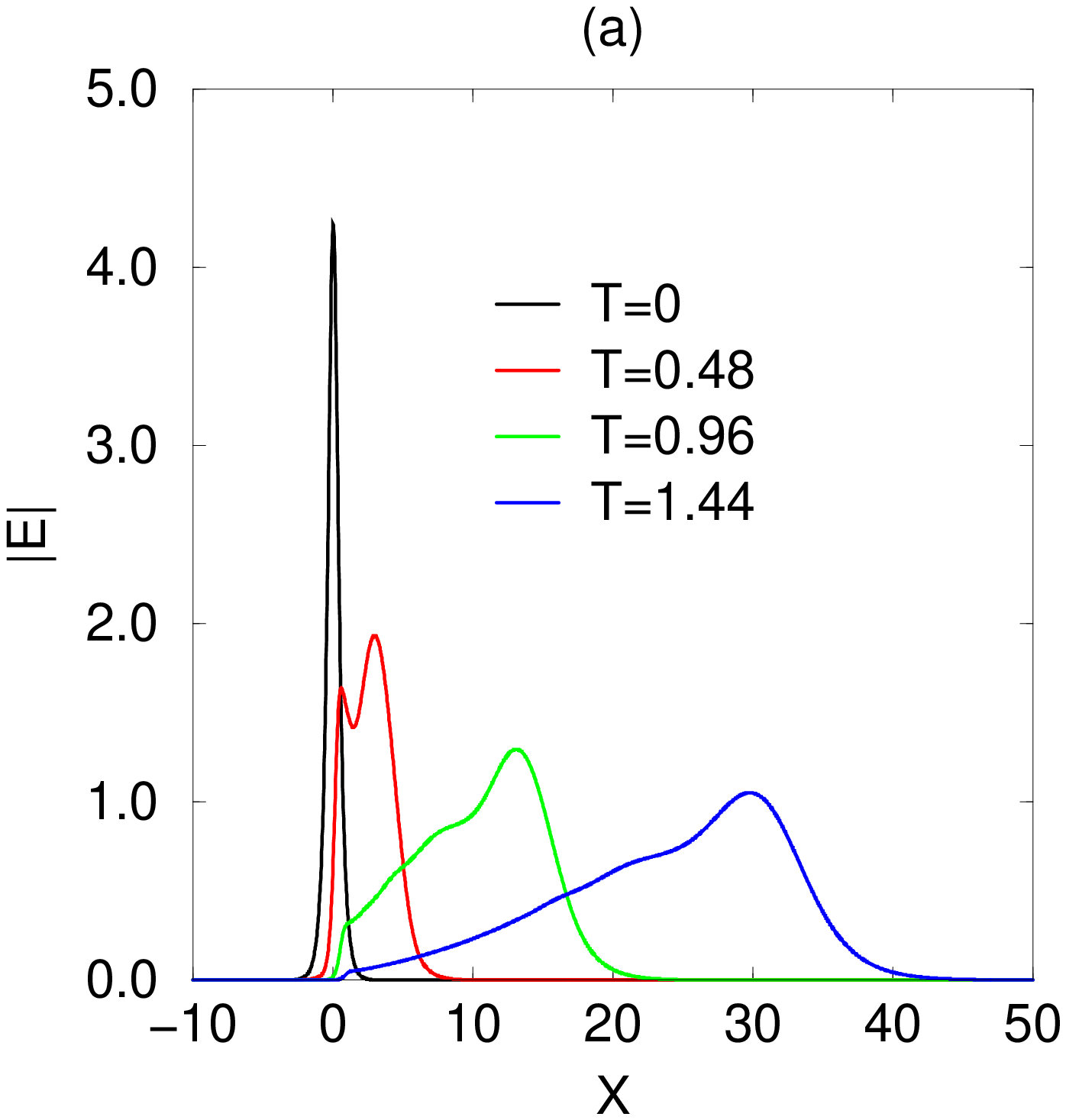}
\includegraphics[height=5.6cm,angle=+00] {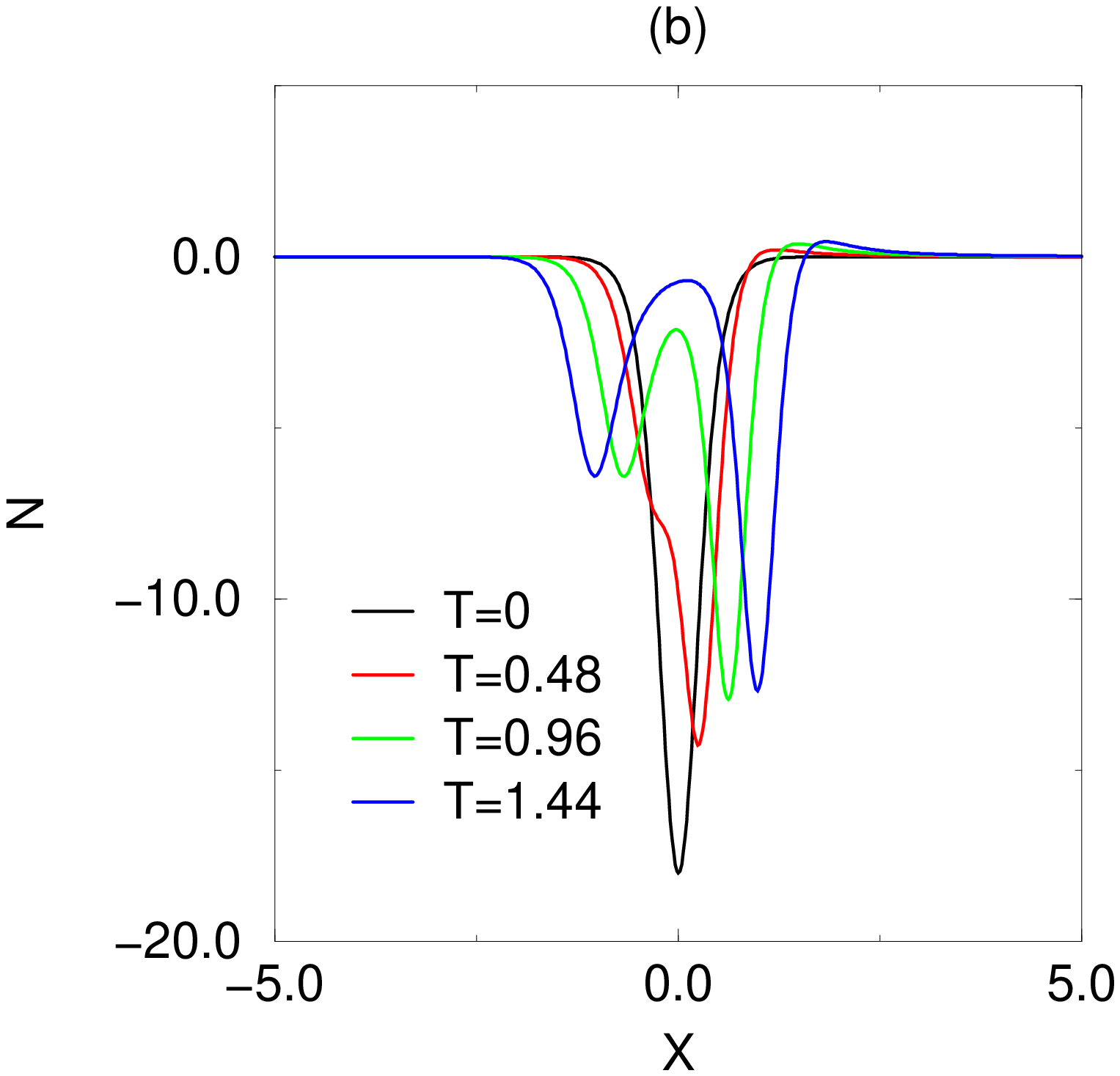}
\caption{(a) Electric field soliton, and (b) density cavity at $L = 50$.
Collapses of the Langmuir soliton caused by a quick escape of the electric field soliton. }
\end{figure}

The collapse of the caviton at the large density gradient limit
can be understood intuitively by comparing the Zakharov system at the adiabatic limit
to the Schr\"{o}dinger equation in quantum mechanics
\begin{eqnarray*}
i \hbar \frac{\partial \psi}{\partial t} 
= -\frac{\hbar^2}{2m}\frac{\partial^2 \psi}{\partial x^2} + V (x) \psi,
\end{eqnarray*}
($\psi$ is the wave function of the quantum system and $\hbar$ is the Planck constant)
which corresponds to a classical Hamiltonian $H (p,x) = p^2/2m + V(x)$ for a point mass $m$, where we
have the correspondence of $H \rightarrow i \hbar \partial_t$ and $p \rightarrow - i \hbar \partial_x$ 
($p$ is the momentum) \cite{sch55}.
From a mathematical analogy, and in comparison with the Hamiltonian, 
the first term in Eqs.(\ref{zak1}) and (\ref{zak4}) corresponds to the kinetic energy
(except for the factor $1/2$ in the first term)
and the terms on the right side of Eqs.(\ref{zak1}) and (\ref{zak4}) correspond to potential; 
$V(x) \leftrightarrow N - \alpha X $.
Without the non-linearity, the point mass (soliton) falls down the potential (density) hill,
toward the right in the cases of Figs.1-3. Figure 4 (a) shows the idea of the soliton motion
versus the potential (which is referred to as a quasi-particle \cite{sag66}). 
In quantum mechanics, because $|\psi|^2 dx$ is proportional to the probability of a point mass found within
a small volume $dx$ \cite{sch55}, when the acceleration is large enough so that the solitons collapse, 
we trace the center of gravity of the $E^2$ profile to determine the particle position.
At each time step, we find a position $X_c$ where $\int_{-\infty}^{X_c} E^2 dX = 1/2 \int_{-\infty}^{\infty} E^2 dX $.
Figure 4 (b) compares the spread of the electric field
in the cases with (solid curves) and without (dashed curves) the nonlinear term in Eq.(3). 
By observing the two cases in Fig.4(b), 
it can be known that the linear potential term
dominates the electric field dynamics when $L=50$. 
Also note that the Langmuir wave eigen-frequency does not exist in
an inhomogeneous plasma due to continuum damping \cite{bar64}.
From a plasma physics point of view,
the acceleration is due to the restoring force in plasma oscillation
being small toward the higher density side.

\begin{figure}[tb]
\centering
  \includegraphics[height=5.6cm,angle=+00] {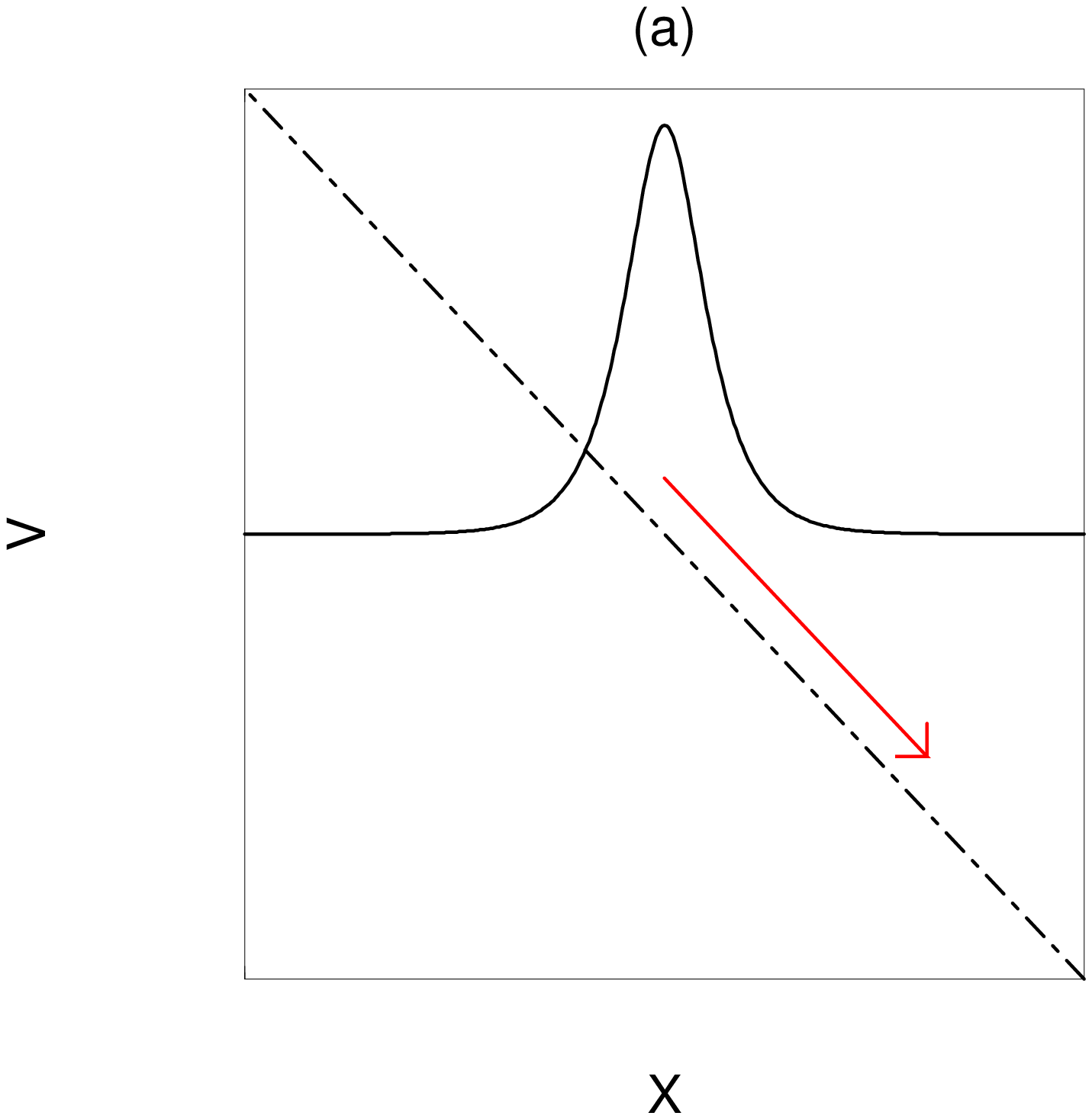}
  \includegraphics[height=5.6cm,angle=+00] {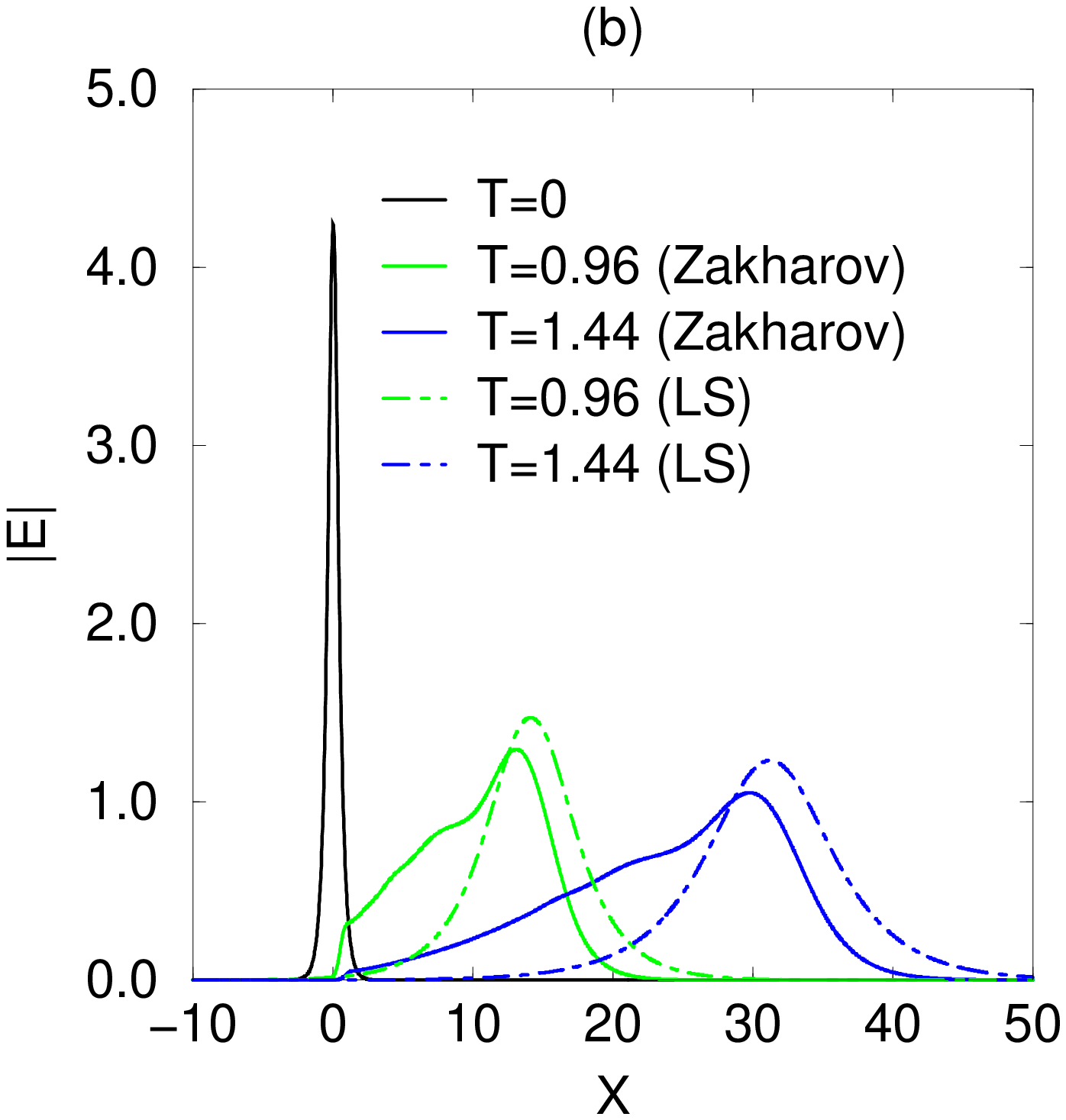}
\caption{Collapse of solitons at high density gradient. (a) Conceptual figure of a soliton
being regarded as a particle falling down toward the right under linearly varying potential.
(b) Spread of the electric field soliton compared
with the solution of a linear Schr\"{o}dinger equation. Here, LS stands for linear
Schr\"{o}dinger.}
\end{figure}

The threshold acceleration (the threshold density gradient) for the caviton collapse is considered.
From the particle motion's analogy, the condition of the kinetic energy being larger
than the potential energy provides us with $V^2/2 = A W \sim N $ as a threshold
($W$ as the soliton width). 
From the potential depth of $N = 2 K_0^2 = 18$ and the soliton width
of $W \simeq 0.5$, we estimate a required acceleration of $A \simeq 36$, or $L=76$, for a complete escape, 
the value of which compares favorably with the threshold obtained in our numerical simulation.
Since the function $sech^2{(K_0 X)}$ has an infinitely long tail,
we set the soliton width at $K_0 X =  1.5$ where the integrated area occupies
$90 \%$ of $\int_{-\infty}^{\infty} E^2 dX$.
Similarly for $K_0=2$, we estimate thresholds for the collapse to be
$L=255$, which also agree favorably with our numerical simulation results.
We understand that the mechanism of the caviton collapse is due to
electric field soliton overcoming the density cavity potential.
On the other hand, if the remaining electric field, or the quasi-particle is still trapped
in the density cavity, it will oscillate within the well.
A signature of the soliton bounce motion (mismatch of the peak positions)
in the potential well is suggested in Fig.5 for a small
density gradient case [obtained from the parameter of Fig.1].

\begin{figure}[tb]
\centering
  \includegraphics[height=5.6cm,angle=+00] {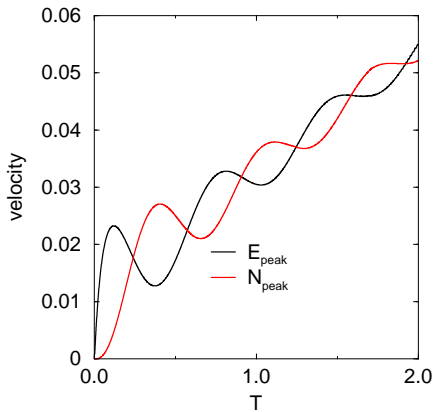}
\caption{Mismatch between the peaks of the electric field soliton and the density cavity for
the $L=5000$ case in Fig.1.}
\end{figure}

Before concluding, a comparison with an adiabatic limit ($N = -|E|^2$) is discussed.
In the Zakharov system, due to the trapping of the electric field by the cavity, 
the acceleration is small compared to the NLSE limit. 
While the acceleration is constant for the NLSE case, much smaller
values of acceleration are found in the Zakharov system.
Taking the $L=500$ case of Fig.2 as example, the acceleration of the Zakharov soliton 
is not constant; rather it is time dependent, as shown in Fig.6,
while $A= 2 \alpha = 5.4$ for the NLSE limit, as predicted theoretically.
This is because the density cavity cannot move faster than the ion sound velocity
and hinder the electric field soliton unless it can completely
escape within the first bounce motion as in the Fig.3 case.
If we solve NLSE with the same parameter and the same initial condition in Eq.(4),
the solitons do not decay simply because they 
do not have emission of the density cavities as in the Zakharov system.

\begin{figure}[tb]
\centering
  \includegraphics[height=5.6cm,angle=+00] {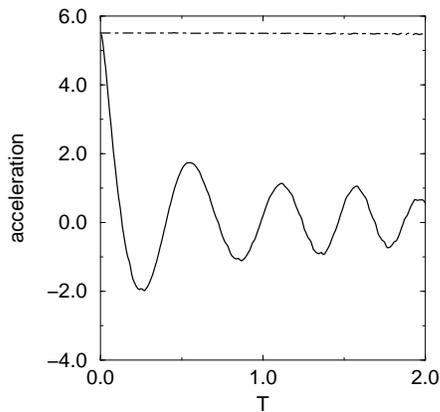}
\caption{The acceleration estimated from the center of gravity for the $L=500$ case in Fig.2.
The solid curve from simulation by the Zakharov equation and the dashed curve by the NLSE.}
\end{figure}

In summary, the Langmuir soliton dynamics in inhomogeneous plasmas was investigated numerically.
By a series of numerical simulations solving Zakharov equations, we have demonstrated
that the solitons are accelerated toward the low density side.
As a consequence of the acceleration and thus a mismatch between the electric field solitons and 
density cavities, isolated cavities moving exactly at the ion sound velocity are emitted.
When the acceleration is further increased, solitons collapse and the cavities 
separate into two lumps released also at the ion sound velocity.
The threshold is estimated by an analogy between the soliton and a particle
overcoming the self-generated potential well.
The current work considered nonlinear wave-wave interactions through Zakharov's fluid model.
In the future, we plan to conduct numerical computation by a Vlasov simulation \cite{czc76,che13}
to incorporate the wave-particle interaction 
to investigate a much more realistic mechanism of 
the Langmuir soliton's sustainment and collapse.

One of the authors (YN) would like to thank discussions with Professor P.K.Kaw.
This work is supported by 
Taiwan NSC 100-2112-M-006-021-MY3 and MOST 103-2112-M-006-007.

\end{document}